\documentclass[%
 reprint,
nofootinbib,
 amsmath,amssymb,
 aps,
prb,
]{revtex4-2}

\usepackage{amsmath,bm}
\usepackage{amssymb}
\usepackage{footnote}
\usepackage{multirow}
\usepackage{graphicx}
\usepackage{subfigure} 
\usepackage{dcolumn}
\usepackage{bm}
\usepackage{bbm}
\usepackage[colorlinks,linkcolor=red,anchorcolor=blue,citecolor=blue]{hyperref}

\newcommand{\tabincell}[2]{\begin{tabular}{@{}#1@{}}#2\end{tabular}}
\newcommand{\Rmnum}[1]{\uppercase\expandafter{\romannumeral #1}}  

\begin{document}

\preprint{APS/123-QED}

\title{Novel method to indirectly  reconstruct  neutrinos  in collider experiments}

\author{Hongrong Qi}
\affiliation{Department of Physics, National Taiwan University, Taipei 10617, Taiwan, ROC}
\author{Paoti Chang}
\affiliation{Department of Physics, National Taiwan University, Taipei 10617, Taiwan, ROC}

\date{\today}

\begin{abstract}

Neutrinos play a crucial role in particle physics, but cannot be tracked in collider experiments. If more than one neutrino is present in a collision event, it is impossible to extract neutrinos' information using any of the traditional methods. In this Letter, we introduce an innovative inclusive-tagging scheme that is capable of capturing the four-momentum of an undetected particle on the signal side, such as a neutrino, $K_L^0$, etc., in collider experiments. 
This is the first proposed solution to the longstanding challenge outlined above. 
Our scheme, based on an asymptotically recursive vector sequence, has the potential to catalyze a significant transformation in the reconstruction techniques for (semi-)leptonic decays of collider experiments. 
The application and development of this scheme will greatly improve the precision of measurements of Standard Model parameters in the (semi-)leptonic sector with the same data samples, and could play a pivotal role in the search for new physics.
Additionally, the asymptotically recursive (vector) sequence introduced in our scheme might also have promising applications in other fields, such as machine learning. 

\end{abstract}
\maketitle


\section{\label{sec:level1}
Introduction}
The primary physics goals of most collider experiments are to improve the precision of measurements of Standard Model (SM) parameters and to search for new physics (NP) beyond SM. 
(Semi-)leptonic decays provide precision tests in the electroweak sector of SM~\cite{BaBar:2014omp,Belle-II:2018jsg}.
Most golden (semi-)leptonic decays involve neutrinos~\cite{Belle-II:2018jsg}, which cannot be detected directly in collider experiments up to the present.

If only one neutrino (or one missing particle\footnote{Throughout this paper, a ``missing" particle refers to a particle that cannot be reconstructed or detected by the tracking system in collider experiments, such as a neutrino or a $K_L^0$.}) is present in an event, its four-momentum can be determined by the recoiling technique~\cite{BESIII:2018gvg}. However, if two or more neutrinos are in an event, all traditional methods become ineffective. In such cases, experimental groups~\cite{BaBar:2014omp,Belle-II:2018jsg,PDG2022} can only extract signal yield from momentum distributions. Both the signal and background in the momentum spectra are shown as (quasi-)smooth distributions, making them difficult to distinguish. The distinction between the two contributions mainly relies on Monte Carlo (MC) simulations. These will result in lower significance and higher uncertainty.
Moreover, the processes related to NP have extremely small branching fractions while the hadronic tagging method has lower efficiency.
To achieve higher efficiency, the tagged side is reconstructed from all remaining tracks and neutral clusters in the event that are not associated with the signal side, a technique known as the inclusive-tagging (or untagged) method~\cite{Belle-II:2018jsg,Likhomanenko:2016tgu}. Its efficiency is approximately 2-3 orders of magnitude higher than that of the hadronic tagging method in $B$ decays.
Although the traditional inclusive-tagging method is a highly efficient reconstruction approach in (semi-)leptonic beauty, charm, and $\tau$  experiments,
its sensitivity is generally worse than that of the hadronic tagging method, mainly attributed to the higher background level and the lower capability to separate signal from background~\cite{Belle-II:2018jsg}.

In this Letter, we introduce an innovative inclusive-tagging scheme capable of capturing the four-momentum of an undetected particle on the signal side, such as a neutrino, $K_L^0$, etc., in collider experiments. 
This is the first time that a solution is introduced to address the aforementioned challenge.
Our scheme, based on an asymptotically recursive vector sequence, has the potential to catalyze a significant transformation in the capabilities of the inclusive tagging method.
When our scheme is further confirmed and developed by future experiments, 
it will greatly improve the precision of measurements of Standard Model parameters in the (semi-)leptonic sector using the same data samples, and will play a pivotal role in the search for NP.
This novel tool can be applied in Belle II, BESIII, LHCb, and other potential collider experiments.
In addition, the asymptotically recursive (vector) sequence introduced in our scheme may also be expected to be a filter in machine learning.

\section{Introduce innovative scheme}
\label{sec:kmatrix}

\begin{figure}
    \vskip 0.4cm
    \centering
    \includegraphics[width=0.45\textwidth]{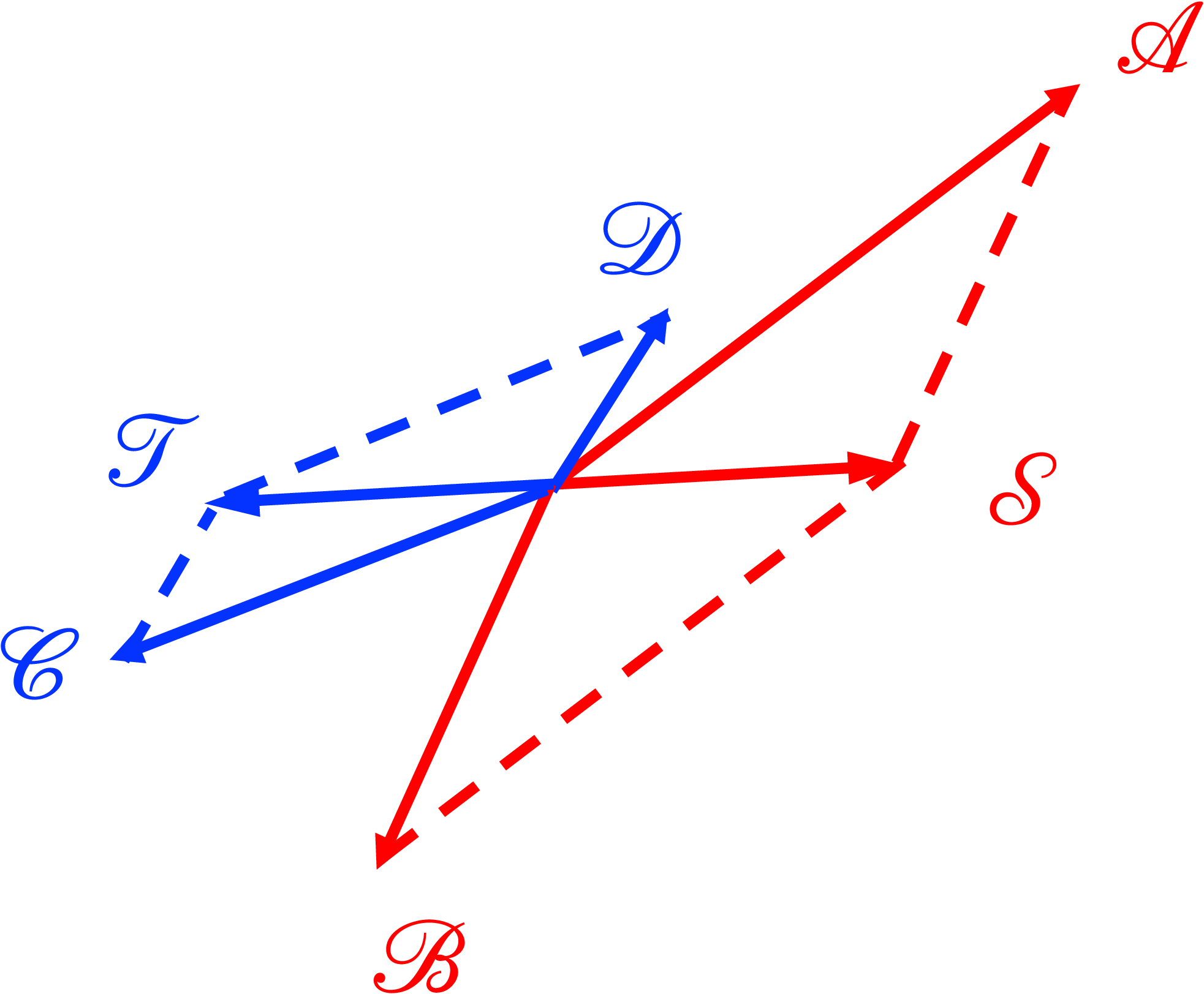}
    \caption{(Color online) Schematic diagram of decays of the $\mathcal{ST}$ system, where the red (blue) parts represent the particles on the signal (tagged) side.}
    \label{fig:decays}
\end{figure}

An event generated by a collider can be commonly classified as signal ($\mathcal{S}$) and tagged ($\mathcal{T}$) sides, where $\mathcal{S}$ and $\mathcal{T}$ originate from a narrow-width particle or are produced directly at the collision point. 
As shown in Fig.~\ref{fig:decays}, the decays are expressed as 
\begin{equation}
\begin{aligned}
\mathcal{S}&\to \mathcal{A} + \mathcal{B} \ ,  \\
\mathcal{T}&\to \mathcal{C} + \mathcal{D} \ ,  \\
\end{aligned}  
\label{eq:decays}
\end{equation}
where, $\mathcal{S}$ and $\mathcal{T}$ refer to narrow-width candidates to be selected, $\mathcal{A}$ denotes the reconstructed part on the signal side, $\mathcal{B}$ is a missing long-lived particle like a neutrino or a $K_L^0$, $\mathcal{C}$ and $\mathcal{D}$ indicate the reconstructed and unreconstructed particles on the tagged sides, respectively. In general, $\mathcal{C}$ 
is known as the rest of the event (ROE) against $\mathcal{S}$.
Since the collider experiments, up to the present, cannot detect neutrinos, the key issue is trying to gain the three-momentum of either $\mathcal{S}$ or $\mathcal{T}$.

In the rest frame of the $\mathcal{ST}$ system,
it has always
\begin{equation}
\begin{aligned}
\bm{p}(\mathcal{B}) + \bm{p}(\mathcal{D})  \equiv -[\bm{p}(\mathcal{A}) + \bm{p}(\mathcal{C})]  \ ,  \\
\end{aligned}  
\label{eq:identity}
\end{equation}
where $\bm{p}(i)$ is the three-momentum vector of the $i$-th term ($i=\mathcal{A}, ~\mathcal{B},~\mathcal{C}$, $\mathcal{D}$), the left and right of the equation represent the undetected and reconstructed parts, respectively.
To obtain the three-momentum of the missing $\mathcal{B}$ particle, the following asymptotically recursive vector sequences can be built in the rest frame of the $\mathcal{ST}$ system,
\begin{equation}
\begin{aligned}
\bm{p}(\mathcal{B})_0&=-\bm{p}(\mathcal{A}) - \bm{p}(\mathcal{C}) \ ,  \\
\bm{p}(\mathcal{B})_1&=\frac{\bm{p}(\mathcal{B})_0+\bm{p}(\mathcal{B})^\prime_0}{2} \\
&=\frac{\bm{p}(\mathcal{B})_0 + [-\bm{p}(\mathcal{A}) - \bm{p}(\mathcal{C}) - \bm{p}(\mathcal{D})_0] }{2} \\
      &=-\bm{p}(\mathcal{A}) - \bm{p}(\mathcal{C}) - \bm{p}(\mathcal{D}) + \bm{p}(\mathcal{D}) - \frac{1}{2}\bm{p}(\mathcal{D})_0 \\
      &\simeq \bm{p}(\mathcal{B}) + \frac{1}{2}\bm{p}(\mathcal{D}) \ ,\\
\vdots & \\
\bm{p}(\mathcal{B})_k&=\frac{\bm{p}(\mathcal{B})_{k-1}+\bm{p}(\mathcal{B})^\prime_{k-1}}{2} \\
&=\frac{\bm{p}(\mathcal{B})_{k-1} + [-\bm{p}(\mathcal{A}) - \bm{p}(\mathcal{C}) - \bm{p}(\mathcal{D})_{k-1}] }{2} \\
 &\simeq\bm{p}(\mathcal{B}) + \frac{1}{2^k}\bm{p}(\mathcal{D}) \ , \\
\vdots & \\
\end{aligned}  
\label{eq:sequence}
\end{equation}
where $k=1,2,3,\cdots$.
Such, the general term $\bm{p}(\mathcal{B})_n$ is extracted as
\begin{widetext}
\begin{equation}
\bm{p}(\mathcal{B})_n=\left\{
\begin{aligned}
&-\bm{p}(\mathcal{A}) - \bm{p}(\mathcal{C}), & &n=0 \ ; \\
&\frac{\bm{p}(\mathcal{B})_{n-1} + \bm{p}(\mathcal{B})_{n-1}^\prime}{2} \simeq \bm{p}(\mathcal{B}) + \frac{1}{2^n}\bm{p}(\mathcal{D}), & &n=1,2,3,\cdots \ ;
\end{aligned}  
\right.
\label{eq:general}
\end{equation}
\end{widetext}
where $\bm{p}(\mathcal{B})_{n-1}^\prime=-\bm{p}(\mathcal{A}) - \bm{p}(\mathcal{C}) - \bm{p}(\mathcal{D})_{n-1}$, and $\bm{p}(\mathcal{D})_{n-1}$ is defined in Eqs.~(\ref{eq:ini}) and (\ref{eq:pD}). The symbols without subscripts, $\bm{p}(\mathcal{B})$ and $\bm{p}(\mathcal{D})$, denote the corresponding truth values; the same applies hereinafter. 

As $n$ becomes infinite,
\begin{equation}
\lim_{n \to \infty } {\bm{p}(\mathcal{B})_n}=\lim_{n \to \infty }{[ \bm{p}(\mathcal{B}) + \frac{1}{2^n}\bm{p}(\mathcal{D}) ]} 
=\bm{p}(\mathcal{B}) \ . 
\label{eq:infty}
\end{equation}
That is, $\bm{p}(\mathcal{B})_n$ is asymptotic to the truth $\bm{p}(\mathcal{B})$.
It is additionally noteworthy that $\bm{p}(\mathcal{D})$ of the missing particle(s) will be ``eaten" by infinite iterations.

It is found that the above sequences strictly hold in mathematics. It is also applicable to four-momentum vectors, since the energies of $\mathcal{A}$, $\mathcal{B}$, $\mathcal{C}$, and $\mathcal{D}$ can be known/determined according to energy-momentum conservation equations for decays satisfying Case~(\ref{eq:decays}) in experiments. However, it is crucial to initialize/parameterize $\bm{p}(\mathcal{D})_k$ ($k=0,1,2,...$) in physics.
Since $\mathcal{S}$ and $\mathcal{B}$ are narrow-width particles, one can constrain their masses to the corresponding mean values according to the Particle Data Group (PDG)~\cite{PDG2022}.
A common initialization/parameterization of $\bm{p}(\mathcal{D})_k$ is described as follows.
\begin{itemize}
    \item 
    For initialization when $k=0$, a random vector $\bm{p}(\mathcal{D})^{\rm rand}$ can be generated, where the magnitude of $\bm{p}(\mathcal{D})^{\rm rand}$ must be less than or equal to the $\mathcal{D}$'s energy. Then, $\bm{p}(\mathcal{S})^{\rm rand}=-\bm{p}(\mathcal{C})-\bm{p}(\mathcal{D})^{\rm rand}$ .
    To obtain the more precisely initial value, one can scale the magnitude of  $\bm{p}(\mathcal{S})^{\rm rand}$ to $\sqrt{E^2(\mathcal{S})-m^2(\mathcal{S})}$, dubbed as $\bm{p}(\mathcal{S})_{0}^{\rm scale}$, where $E(\mathcal{S})$ and $m(\mathcal{S})$ are the energy and mass of $\mathcal{S}$, respectively. We define $\bm{p}(\mathcal{B})^\alpha_0=\bm{p}(\mathcal{S})_0^{\rm scale}-\bm{p}(\mathcal{A})$, and then constrain the invariant mass of $\mathcal{B}$ to its mean mass~\cite{PDG2022}, denoted as $\bm{p}(\mathcal{B})^{\beta}_0$. Thus, 
    \begin{equation}
    \label{eq:ini}
        \bm{p}(\mathcal{D})_0 = -\bm{p}(\mathcal{A}) - \bm{p}(\mathcal{C}) - \bm{p}(\mathcal{B})^{\beta}_0 \ .
    \end{equation}
    
    \item For parameterization when $k=1,2,3,...$, $\bm{p}(\mathcal{D})_k$ can be got from the two following methods:
    \begin{equation}
    \begin{aligned}
     \bm{p}(\mathcal{D})^{\rm \Rmnum{1}}_k &= - \bm{p}(\mathcal{S})_{k-1}^{\rm scale} - \bm{p}(\mathcal{C}) \ , \\
     \bm{p}(\mathcal{D})^{\rm \Rmnum{2}}_k &= -\bm{p}(\mathcal{A}) - \bm{p}(\mathcal{C}) - \bm{p}(\mathcal{B})^{\rm \Rmnum{1}}_{k-1}  \ ,
    \end{aligned}  
    \label{eq:par}
    \end{equation}
    
    where $\bm{p}(\mathcal{S})_{k-1}^{\rm scale}$ is obtained by scaling the magnitude of  $\bm{p}(\mathcal{A}) + \bm{p}(\mathcal{B})_{k-1}$ to $\sqrt{E^2(\mathcal{S})-m^2(\mathcal{S})}$, 
    and $\bm{p}(\mathcal{B})^{\rm \Rmnum{1}}_{k-1}= \bm{p}(\mathcal{B})_{k-1} - \frac{1}{2^k}  \bm{p}(\mathcal{D})^{\rm \Rmnum{1}}_{k-1}$ . 
    Carrying Eq.~(\ref{eq:general}) into $\bm{p}(\mathcal{B})^{\rm \Rmnum{1}}_{k-1}$, it is easy to find,
    \begin{widetext}
    \begin{equation}
    \begin{aligned}
    \lim_{k \to \infty }{\bm{p}(\mathcal{D})^{\rm \Rmnum{2}}_k}
    &= -\bm{p}(\mathcal{A}) - \bm{p}(\mathcal{C}) - \lim_{k \to \infty }{\bm{p}(\mathcal{B})^{\rm \Rmnum{1}}_{k-1}} \\
    &= -\bm{p}(\mathcal{A}) - \bm{p}(\mathcal{C}) 
     - \lim_{k \to \infty }{\left[\bm{p}(\mathcal{B}) 
     + \frac{1}{2^{k-1}} \bm{p}(\mathcal{D}) - \frac{1}{2^k}  \bm{p}(\mathcal{D})^{\rm \Rmnum{1}}_{k-1} \right] }   \\
    &= -\bm{p}(\mathcal{A}) - \bm{p}(\mathcal{C}) - \bm{p}(\mathcal{B}) + \lim_{k \to \infty }{\left[\frac{1}{2^{k-1}} \left(\bm{p}(\mathcal{D}) - \frac{1}{2}  \bm{p}(\mathcal{D})^{\rm \Rmnum{1}}_{k-1} \right) \right] } \\
    &= \bm{p}(\mathcal{D})~.
    \end{aligned}
    \label{eq:pD2}
    \end{equation}
    \end{widetext}
    
    Finally, the arithmetic mean of $\bm{p}(\mathcal{D})^{\rm \Rmnum{1}}_k$ and $\bm{p}(\mathcal{D})^{\rm \Rmnum{2}}_k$ is taken as $\bm{p}(\mathcal{D})_k$, i.e. 
    \begin{equation}
    \bm{p}(\mathcal{D})_k = \frac{\bm{p}(\mathcal{D})^{\rm \Rmnum{1}}_k + \bm{p}(\mathcal{D})^{\rm \Rmnum{2}}_k}{2} \ .
    \label{eq:pD}
    \end{equation}
\end{itemize}

As for iteration times, $k=15$ is generally enough as $1/2^{15}=1/32768<0.01\%$.
    
It should be noted that other approaches for initializing/parameterizing $\bm{p}(\mathcal{D})_k$ that satisfy users' requirements and have no bias peaks are also desirable.

\section{\label{sec:level3} Test in pseudo-experiment data}

Three pseudo-experiment samples are generated with ROOT software and used to test the scheme we proposed in the previous section. The following channels are carried out.

\begin{itemize}
    \item $\Upsilon(4S)\to B^+ B^- \to \mu^+\nu_{\mu} + X$ which is applicable to the Belle II experiment. (Throughout this paper, $X$ represents the fully final states on the tagged side, i.e. $X=\mathcal{T}= \mathcal{C} + \mathcal{D}$, and the charge-conjugate mode is implicitly included; this convention is maintained unless explicitly indicated otherwise.)
    \item $B^0 \to \tau^-\tau^+ \to \pi^-\pi^+\pi^-\nu_{\tau} + X$,  applicable to the Belle II and LHCb experiments.
    \item $e^+e^- \to  (\gamma_{\rm ISR}) \Lambda_c^+ \Lambda_c^- \to (\gamma_{\rm ISR})  \Lambda e^+ \nu_{e} + X$, applicable to the Belle II and BESIII experiments.
\end{itemize}

\begin{figure*}[htb]
\centering
\includegraphics[width=0.35\textwidth]{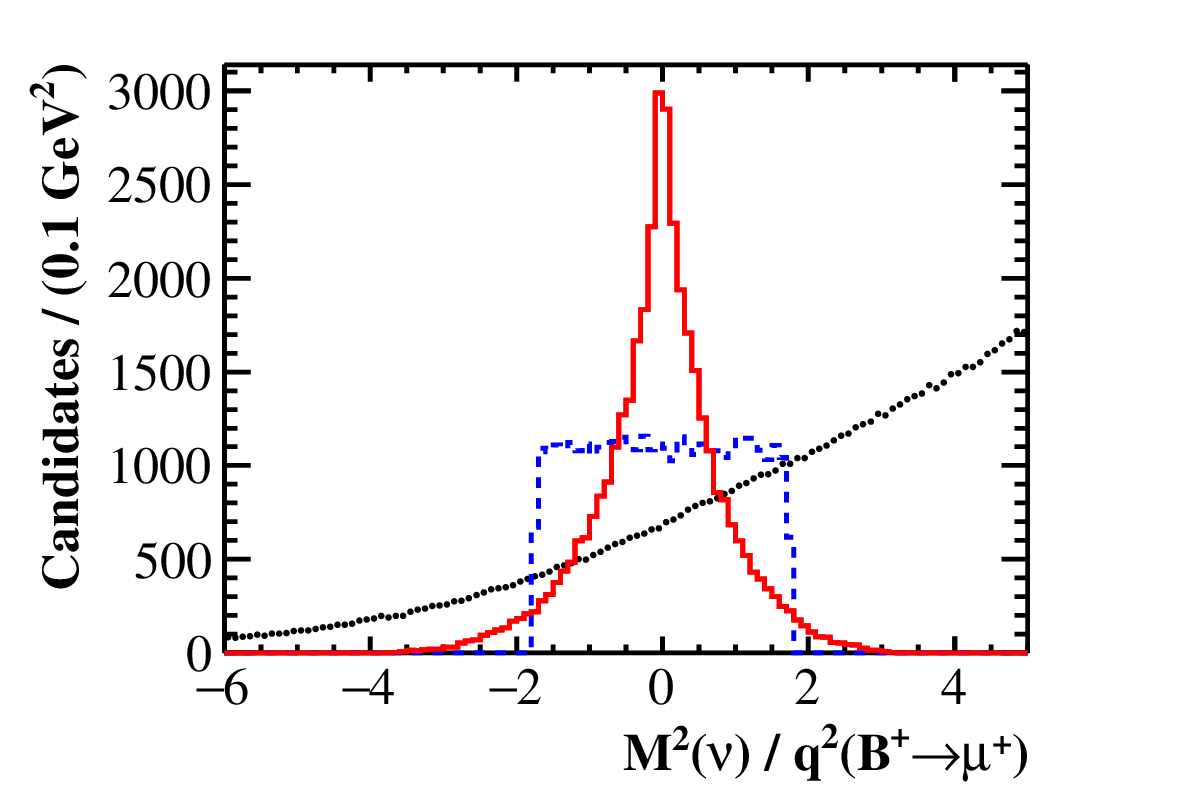} 
\includegraphics[width=0.35\textwidth]{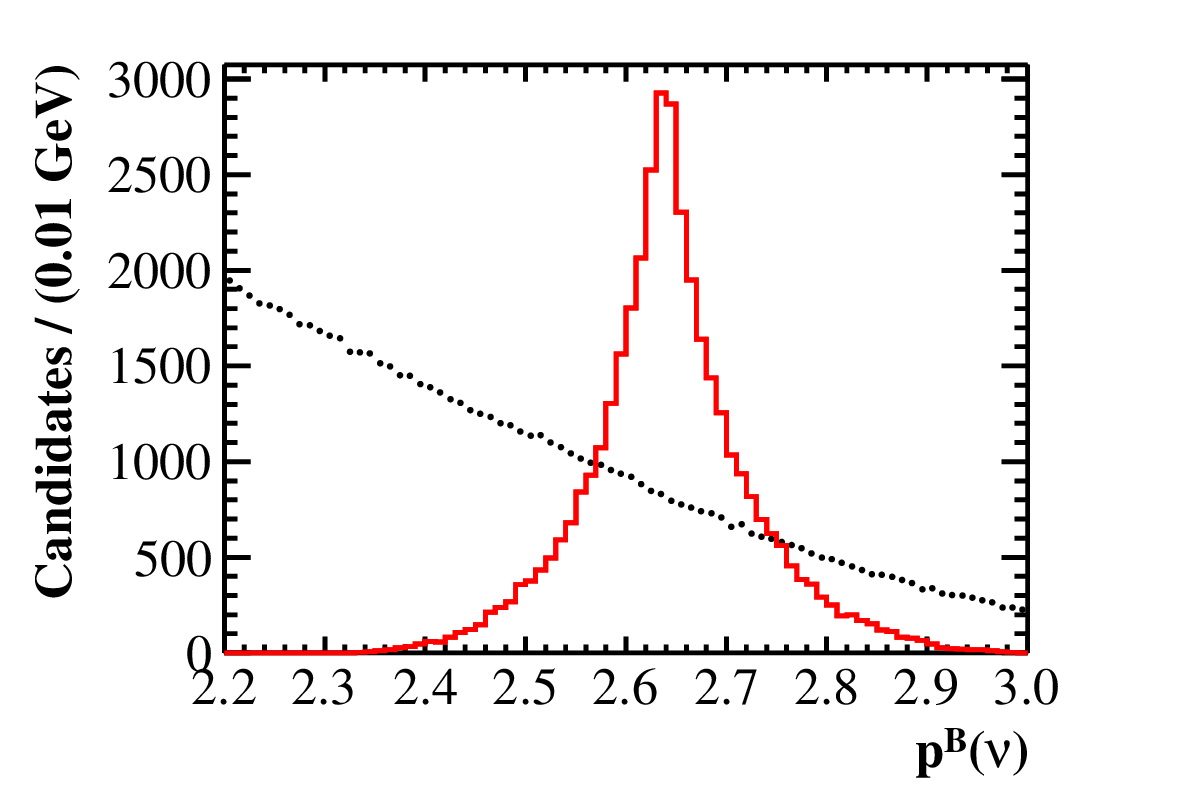} 
\caption{(Color online) Test of $\Upsilon(4S)\to B^+ B^- \to \mu^+\nu_{\mu} + X$  in the pseudo-experiment data. In the left subfigure, the solid red line represents the mass squared $M^2$ [when the ``detected" momentum squared $p^2$ is less than the energy squared $E^2$, there is $M^2=-\sqrt{-(E^2-p^2)}$; the same applies hereinafter], the dashed blue line describes the quasi four-momentum transfer squared $q^2$, and the dotted black is the mass-squared spectrum on an arbitrary scale from the continuum background pseudo-experiment data. In the right subfigure, the solid red line shows the momentum distribution of neutrinos in the $B^+$ rest frame, and the dotted black line is the corresponding contribution on an arbitrary scale from the continuum background pseudo-experiment data.}
\label{fig:toy:b2munu}
\end{figure*} 
\begin{figure*}[htb]
\centering
\includegraphics[width=0.35\textwidth]{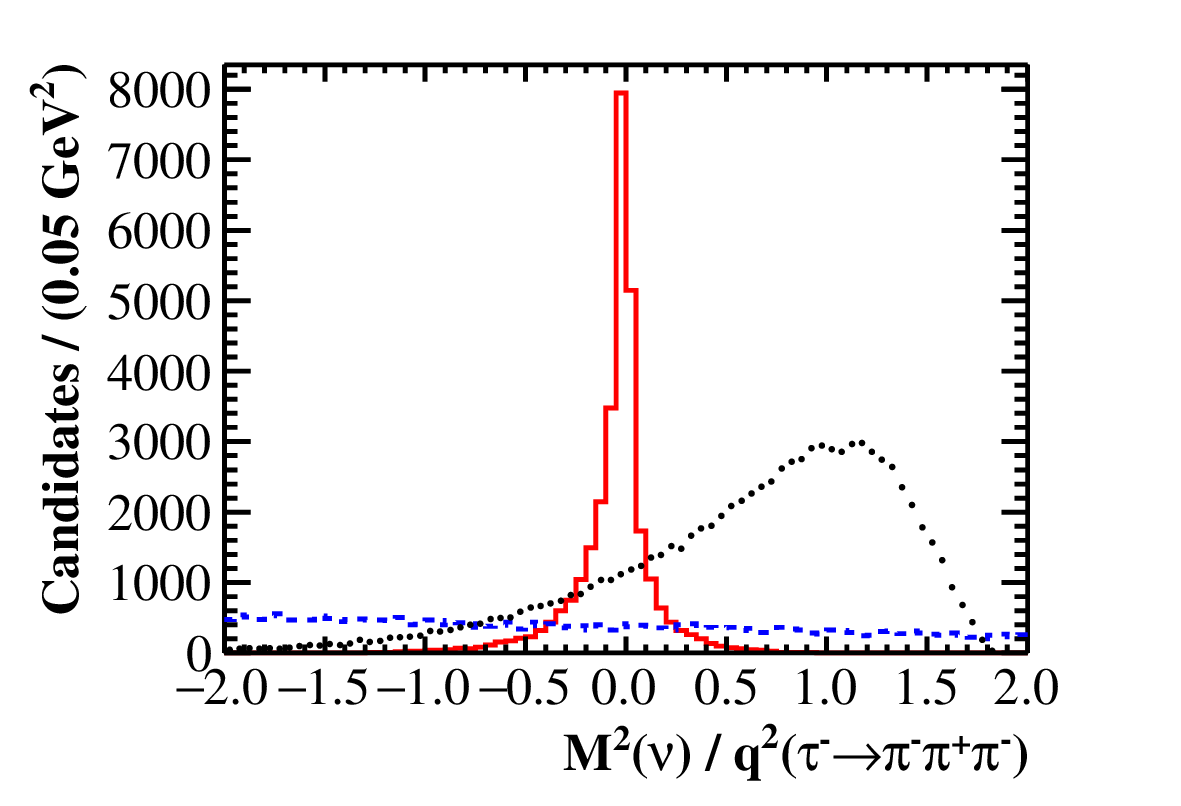}
\includegraphics[width=0.35\textwidth]{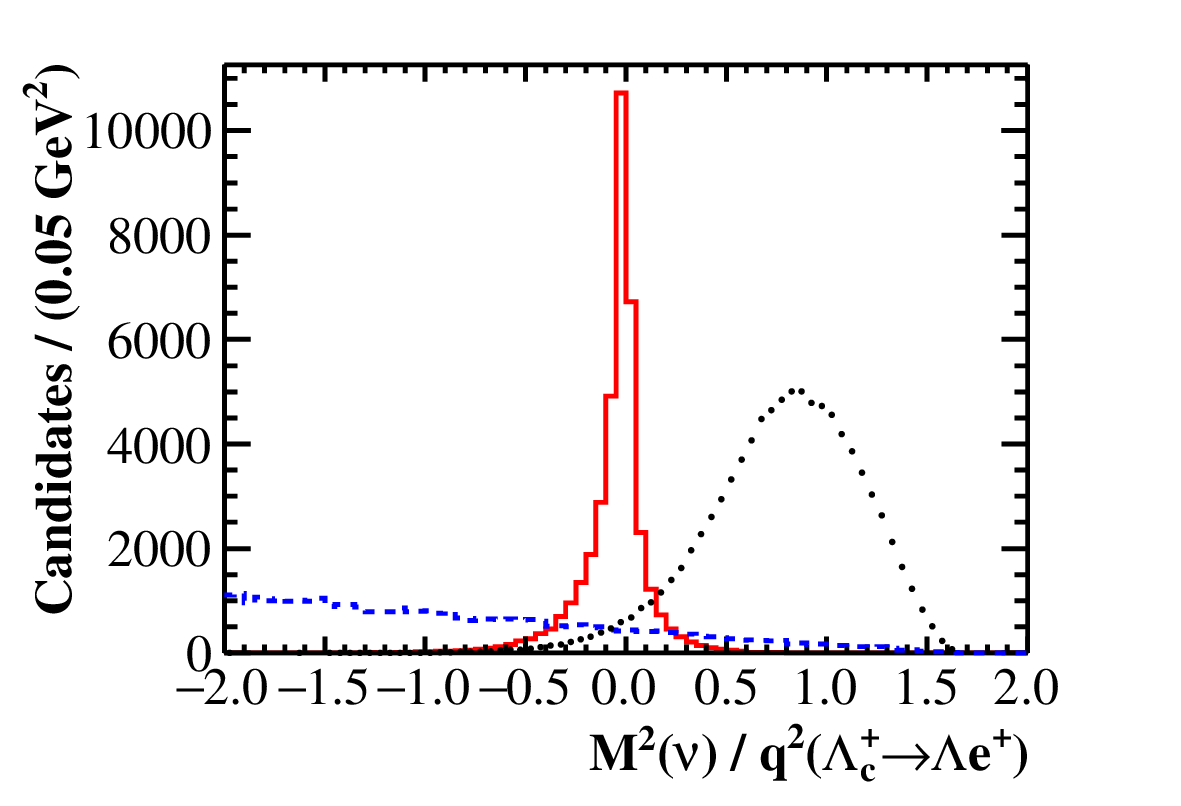}  
\caption{(Color online ) Test of $B^0 \to \tau^-\tau^+ \to \pi^-\pi^+\pi^-\nu_{\tau} + X$ (left) and $e^+e^- \to (\gamma_{\rm ISR})  \Lambda_c^+ \Lambda_c^- \to (\gamma_{\rm ISR})  \Lambda e^+ \nu_e + X$  at $\sqrt{s}=5.0$ GeV (right) in the pseudo-experiment data. These lines have meanings similar to those on the left of Fig.~\ref{fig:toy:b2munu}. }
\label{fig:toy:others}
\end{figure*}

In simulations, $\mathcal{A}$ (or $\mathcal{C}$) is regarded as a whole part when $\mathcal{A}$ (or $\mathcal{C}$) is not a single particle, which can be approximated as a ``virtual" particle that has a variant mass spectrum and a huge-wide width.
All decays are generated by the Phase Space function with ROOT::TGenPhaseSpace class~\cite{Root}.
The resolution is set to be $\sigma_P/P=0.5\%$ for one track on the signal side, while $\sigma_P/P=1.0\%$ for ROE since the ROE is a combination of several particles.
Differences in efficiency for each track/part are negligible, which is expected to have few impacts on this scheme.
The common requirement for ROE is implemented in the same way as described in Ref.~\cite{Belle-II:2023esi}, $E(\mathcal{C})\in [\frac{1}{2} E(\mathcal{S}), E(\mathcal{S}) + 0.5]$~GeV.

In general, the invariant mass squared of a neutrino $M^2(\nu)$  or 
the quasi four-momentum transfer squared $q^2=E(\mathcal{S})^2-2\times E(\mathcal{S})\times E(\mathcal{A})+m^2(\mathcal{A})$ (generally assuming $\mathcal{S}$ to be stationary in the rest frame of the $\mathcal{ST}$ system in the traditional inclusive-tagging method) are used to express the mass-related term in connection with neutrinos~\cite{Belle-II:2023esi}, where $m^2(\mu)$ is the mean muon's mass~\cite{PDG2022}.
As shown in Figs.~\ref{fig:toy:b2munu} and \ref{fig:toy:others}, the mean neutrino's masses in the above three modes are all around zero. In contrast, the traditional $q^2$ in each mode is shown as a distribution, which is difficult to distinguish from backgrounds.
For the $B^+ \to \mu^+\nu_{\mu}$ process in the first pseudo-experiment sample, the muon and the neutrino flight back to back in the $B^+$ rest frame. The momentum of each daughter in the $B^+$ rest frame is calculated to be 2.639 GeV/c according to the PDG's parameters~\cite{PDG2022}.
The fit to the momentum distribution of neutrinos (the right subfigure of Fig.~\ref{fig:toy:b2munu}) gets $p^B_{\nu}=2.6391\pm0.0004$ GeV (natural units with $\hbar=c=1$ is used throughout this Letter), consistent with the above calculation.
In addition, the continuum backgrounds are also generated and tested. As shown in Figs.~\ref{fig:toy:b2munu} and \ref{fig:toy:others}, no bias peaks are found. 
These confirm that the innovative scheme is effective and reliable.

\section{\label{sec:level4} Summary and perspective}
We innovatively proposed an inclusive-tagging scheme that can obtain the four-momentum of an undetected long-lived particle in collider experiments, such as Belle-II, BESIII, LHCb, and other potential experiments. This scheme, based on an asymptotically recursive (vector) sequence, strictly holds in mathematics. From the pseudo-experiment data, it is proved to be effective and reasonable in experimental particle physics.
It can be expected to be further confirmed and applied by future experiments.
Besides,  the novel idea of this scheme,
the asymptotically recursive (vector) sequences, might serve as one of the filters in machine learning so that some missing/unknown information will be ``eaten" by infinite iterations (which will be the focus of our future research).

\begin{acknowledgments}
We are deeply grateful for Yuan Chao's contributions, including some discussions and a few of tests. We thank the support of Grants
No. MOST 110-2639-M-002-002-ASP, NSTC 112-2639-M-002-006-ASP
and  MOST 110-2112-M-002-022-MY3.
\end{acknowledgments}

\appendix*
\section{Supplementary Notes}

\begin{table*}
\caption{\label{tab:app}Additional remarks on the three pseudo-experiment cases, here $X$ denotes the fully final states on the tagged side.}
\begin{ruledtabular}
\begin{tabular}{ccccc}
\noalign{\smallskip}
Case & $\mathcal{A}$ & $\mathcal{B}$ & $\mathcal{C}$ & $\mathcal{D}$ \\ \noalign{\smallskip}\hline\noalign{\smallskip}

$\Upsilon(4S)\to B^+ B^- \to \mu^+\nu_{\mu} + X$ & $\mu^+$ & $\nu_{\mu}$ &  \tabincell{c}{ reconstructed part on \\ the tagged $B^-$ side} & \tabincell{c}{unreconstructed part on \\ the tagged $B^-$ side} \\ \noalign{\smallskip}\hline\noalign{\smallskip}

$B^0 \to \tau^-\tau^+ \to \pi^-\pi^+\pi^-\nu_{\tau} + X$ & $\pi^-\pi^+\pi^-$ & $\nu_{\tau}$ & \tabincell{c}{ reconstructed part on \\ the tagged $\tau^+$ side} & \tabincell{c}{unreconstructed part on \\ the tagged $\tau^+$ side} \\ \noalign{\smallskip}\hline\noalign{\smallskip}

$e^+e^- \to  (\gamma_{\rm ISR}) \Lambda_c^+ \Lambda_c^- \to (\gamma_{\rm ISR})  \Lambda e^+ \nu_{e} + X$ & $\Lambda e^+$ &  $\nu_{e}$ & \tabincell{c}{ reconstructed part on \\ the tagged $\Lambda_c^-$ side} & \tabincell{c}{unreconstructed part on \\ the tagged $\Lambda_c^-$ side} \\ \noalign{\smallskip} 
\end{tabular}
\end{ruledtabular}
\end{table*}

\bibliography{main}

\end{document}